# Probing weak force induced parity violation by high resolution mid-infrared molecular spectroscopy

S. K. Tokunaga[1,2], C. Stoeffler[1,2,†], F. Auguste[1,2], A. Shelkovnikov[1,2,‡], C. Daussy[1,2], A. Amy-Klein[1,2], C. Chardonnet[2,1]\*, B. Darquié[2,1]

[1] *Université Paris 13, Sorbonne Paris Cité, Laboratoire de Physique des Lasers, F-93430, Villetaneuse, France*

[2] *CNRS, UMR 7538, LPL, F-93430, Villetaneuse, France.*

\* chardonnet@univ-paris13.fr

† Present address: Groupe de spectrométrie moléculaire et atmosphérique (GSMA), CNRS : UMR7331 – Université de Reims - Champagne Ardenne, France

‡ Permanent address: P.N. Lebedev Physics Institute, Russian Academy of Sciences, Leninsky prosp. 53, 119991 Moscow, Russia

**Abstract**

To date no experiment has reached the level of sensitivity required to observe weak nuclear force induced parity violation (PV) energy differences in chiral molecules. In this paper, we present the approach, adopted at Laboratoire de Physique des Lasers (LPL), to measure frequency differences in the vibrational spectrum of enantiomers. We review different spectroscopic methods developed at LPL leading to the highest resolutions, as well as 20 years of $CO_2$ laser stabilization work enabling such precise measurements. After a first attempt to observe PV vibrational frequency shifts using sub-Doppler saturated absorption spectroscopy in a cell, we are currently aiming at an experiment based on Doppler-free two-photon Ramsey interferometry on a supersonic beam. We report on our latest progress towards observing PV with chiral organo-metallic complexes containing a heavy rhenium atom.

**Introduction**

With their rich electronic, vibrational, rotational and hyperfine structure, molecules can play a decisive role in precision tests of fundamental physics. They are now being (or have recently been) used to test fundamental symmetries such as parity[1-3], or both parity and time reversal[4], to test the Pauli exclusion principle[5], to measure either values of fundamental constants (electron-to-proton mass ratio[6], Boltzmann constant[7]), or to measure their variation in time (fine structure constant[8], electron-to-proton mass ratio[9,10]). Most of those experiments can be cast as the measurement of molecular frequencies highlighting the importance of high-resolution molecular spectroscopy and frequency metrology.

In this context, looking for weak nuclear force-induced parity violating effects has been a long-standing quest in molecular physics. The parity operation (P) is one of the three fundamental symmetries along with charge conjugation (C) and time reversal (T) which make up the CPT theorem. Although the theorem states that, in the frame of Lorentz invariant quantum field theories, the combination of the three symmetries must be conserved by all physical processes and by the four fundamental forces (gravitation, the electromagnetic force, the strong and the weak force), it does not exclude that individual ones may be broken. In 1956, Lee and Yang proposed that parity may not be conserved in weak interactions[11]. This proposal was subsequently confirmed by experiment in nuclear[12] and atomic physics (see references 13 and references therein). It has yet to be observed in the molecular domain, where a tiny energy difference between enantiomers of chiral molecules would constitute a unambiguous signature of parity violating electroweak interactions as initially suggested in references 14 and 15. In biology, chirality is a hallmark of life. Nature shows, with very few exceptions, a distinct preference for L-amino acids and D-sugars over their mirror images. The origin of biochirality remains unexplained and there has been speculation that the very tiny energy difference between enantiomers could have provided the bias needed to seed the observed handedness of biological molecules[14, 16]. A successful parity violation (PV) measurement in chiral molecules is thus a test of the Standard Model of Physics in the low-energy regime, and its comparison to quantum chemistry calculations will serve to stimulate further research on the origin of biomolecular homochirality.

To date, Martin Quack's group and our group at Laboratoire de Physique des Lasers (LPL) are among the very few groups to have made experimental attempts to measure PV in chiral molecules. The two groups have chosen different but complementary approaches. However, no experiment to date has reached the required level of resolution needed to observe the tiny PV energy differences involved.

This paper presents the route followed by the LPL group towards the observation of parity violation in molecules. Our approach consists in looking for frequency differences in the spectrum of enantiomers. In doing so, we favour probing the – usually mid-infrared – vibrational spectrum for two reasons. The natural linewidths are narrow (of the order of 1 Hz). And according to Lethokov's rule of thumb[17] for PV shifts exhibited by transitions within a single electronic potential, the magnitude of the PV shift is a fraction of the overall electronic energy shift roughly proportional to the transition's frequency. This leads to a good compromise between the size of the effect and the potentially reachable experimental resolution. But even in the most favourable molecules, the effect remains small; it has been calculated to be of the order of or below 1 Hz. There is thus no hope of reaching the level of resolution required to measure such a minute energy difference in a racemic mixture. The spectra of the two enantiomers must be individually recorded and compared. The sensitivity to PV is then determined by the precision with which any difference between the two measured line centres can be distinguished. Since many systematic effects are proportional to the spectral linewidth, it is crucial to reach the highest possible resolution. However, as detailed below, increasing the resolution often results in decreasing the signal amplitude and in turn the statistical uncertainty on the line centre. Thus a compromise must be found.

Projects aiming at resolving major scientific issues such as observing PV in chiral molecules are inevitably long-term and most often require tremendous experimental and instrumental developments to push both concepts and technologies to their limit. The remainder of the paper is divided in three parts: the first one recalls the different spectroscopic methods and illustrates the path to attaining the highest resolutions. The second gives a detailed historical narrative on the development of high resolution spectrometers in the 10 µm region. Finally, the third recounts our attempts made thus far to measure PV in chiral molecules, and presents our latest experimental efforts towards this goal.

**Molecular ultra-high resolution spectroscopy**

The choice of a specific spectroscopic method for addressing a given scientific objective is guided by the level of resolution required and the need for sufficient signal-to-noise ratio. In the context of the search for frequency differences between right- and left-handed chiral molecules, as already mentioned, the expected shift is so small that there is no experimental method than can resolve the right- and left-handed molecular resonances by probing a racemic mixture. Spectra of both enantiopure samples must be recorded separately and the sensitivity of the experiment is given by the uncertainty on the determined line centre frequency difference. This uncertainty is of the order of $\frac{\Delta}{S/N}$ with $\Delta$ the resonance line width (the resolution) and $S/N$ the signal-to-noise ratio, not limited by the resolution. Although only strictly true if the uncertainty were purely statistical, it remains a good rule of thumb: many systematic uncertainties, a main concern for such high precision experiments, are usually proportional to the resolution. As a consequence, both high resolution and large signal-to-noise ratios are really desirable. In this section, we will compare various ultra-high resolution spectroscopic methods developed at LPL. Rather than quantitatively evaluating signal amplitudes and widths, the aim is to give insight into the underlying physics, the associated orders of magnitude and scaling laws. In doing so, we will illustrate our scientific choices, in particular why we decided to change from sub-Doppler saturated absorption spectroscopy to Doppler-free two-photon Ramsey interferometry for the observation of vibrational frequency differences.

*Linear absorption Doppler-broadened spectroscopy*

The invention of the laser more than fifty years ago opened the route of high resolution spectroscopy since the light source for the first time did not limit anymore the resolution in atomic or molecular spectroscopy (see Section Laser frequency control). Let us only consider the gas phase. In linear absorption spectroscopy, particles' thermal motion limits the resolution to the Doppler width: $\Delta_{\text{Doppler}} = ku = \frac{2\pi}{\lambda}\sqrt{\frac{2k_B T}{m}}$ (e-fold half-width of the Gaussian Doppler profile). $k$ is the modulus of the wave vector $\vec{k}$, $\lambda$ the laser wavelength, $u$ the most probable speed, $k_B$ the Boltzmann constant, *T* the temperature and *m* the mass of the particle under investigation. The resonance condition for a particle of velocity $\vec{v}$ is reached when $\omega_0 = \omega - \vec{k} \cdot \vec{v}$, with $\omega_0/2\pi$ and $\omega/2\pi$ the particle and laser frequency respectively. This resonance condition is relaxed due to the finite coherent particle-light interaction time, limited to $\tau_c = 1/\gamma$, the mean time between two collisions (inverse of the collisional width $\gamma$), or to the transit time in the laser beam $\tau_t \sim w_0/v_\perp$ (with $w_0$ the waist of the laser beam and $v_\perp$ the transverse velocity[a])[18]. For an ensemble of particles in usual conditions of laser power, the relevant transit time is $\tau_t \sim w_0/u$ where *u* is the most probable speed. The resonance condition is also relaxed due to the finite time $\tau_R = \pi/\Omega_R$ required for the particle to absorb a photon (associated to the so-called saturation broadening), with $\Omega_R/2\pi = \dfrac{\mu E_0}{2h}$ the Rabi frequency, $\mu$ the electric dipole moment of the transition and $E_0$ the laser field amplitude. Let us, in a simplified picture, call $\tau$ the smallest of these three times. A particle of given velocity then contributes to the signal over a frequency range equal to $1/\pi\tau$, the full homogeneous width. Reciprocally, particles contributing to the signal at a given frequency belong to a longitudinal (along $\vec{k}$) velocity class of width $\lambda/\pi\tau$. The linear absorption signal amplitude is thus proportional to the homogeneous to Doppler width ratio.

*Saturated absorption spectroscopy*

Various methods were then developed to overcome the Doppler limit and to enter into the domain of very high resolution spectroscopy. The most widely used method is probably saturated absorption spectroscopy for which the gas interacts with a standing wave. The two counterpropagating waves are simultaneously in resonance with a particle when $\omega_0 = \omega - \vec{k} \cdot \vec{v} = \omega + \vec{k} \cdot \vec{v}$. This leads to a narrow resonance centred at $\omega_0 = \omega$ of half width at half maximum equal to $1/\tau$ (following a similar argumentation as in the previous section). Particles contributing to this signal thus cross the laser perpendicularly and have a zero longitudinal velocity with a tolerance of $\pm \lambda/2\pi\tau$. The saturation signal is a Lamb dip in the centre of the Doppler profile[19]. The size of the signal strongly depends on the laser power with an optimum at $\Omega_R \tau = \pi/2$ (corresponding to a so-called $\pi/2$ pulse). However, the amplitude of the saturation signal being a fraction of the Doppler signal, itself proportional to the resolution $1/\tau$, one concludes that the higher the resolution, the smaller the amplitude of the saturated absorption signal.

---

[a] The transverse velocity is defined as the velocity component which is perpendicular to the laser beam.

The time $\tau$ is thus the key parameter when very high resolution is needed. For spectroscopy in the mid-infrared domain or longer wavelengths, the natural life time is usually of the order of 1 s and is not a limitation. The collisional broadening can also be made negligible by working at low enough pressures. The price to be paid is to use a large enough absorption cell to compensate for the reduction of the gas density by an increase in optical length. The resolution is then limited by the transit-time $\sim w_0/u$ through the laser beam. It can indeed be viewed as the consequence of a residual Doppler shift $u\Delta k_\perp$, with $\Delta k_\perp \sim 1/w_0$, the laser beam transverse wave vectors distribution width. Note that this distribution is related to the waist size $w_0$ and not to the beam radius (even if the phrase transit-time is often used) and therefore does not depend on the location along the beam path. In these conditions it is thus important to have a well-collimated beam in order to reach the highest resolutions. As an example, in our group an 18-m long triple path cell was built leading to 108 m of absorption length after retroreflection. We were able to operate in the $10^{-2}$-$10^{-4}$ Pa pressure range and with a 3.5-cm waist leading to a transit-time limited resolution below 1 kHz in the 10-µm spectral region. This setup made possible ultra-high resolution spectroscopy at room temperature of molecules like $SF_6$[20] or CHFClBr, the common chiral prototype molecule[21].

*Two-photon Doppler-free spectroscopy*

An alternative way to overcome the Doppler effect is the Doppler-free two-photon spectroscopy method for which particles interact with a standing wave as well. Consider a three-level (a, b and c) system with intermediate state b connected to a and c by dipole moments. One looks for signals resulting from the absorption of one photon from each laser beam in order to address the c-a transition of frequency $\omega_0$. This can only happen on resonance $\omega_0 = \omega - \vec{k}\cdot\vec{v} + \omega + \vec{k}\cdot\vec{v} = 2\omega$ as the Doppler-shift relative to one beam is exactly compensated by the Doppler-shift relative to the other. The main advantage of 2-photon transitions over saturated absorption is that all particles contribute to the signal, whatever their speed. This is a considerable gain in number of particles equal to the ratio between the Doppler and homogeneous width $\sim ku\tau$ which is, in the context of high resolution spectroscopy, a very large number. The disadvantage is that such transitions often require high laser intensities in order to be driven efficiently (the intermediate state b is often off-resonant for all velocity classes), which results in a larger laser intensity noise and potentially higher systematic effects. The Doppler-free two-photon interaction can be described by introducing an effective Rabi frequency coupling states a and c and proportional to the two electric fields of the standing wave. It can thus be seen as an effective linear absorption spectroscopy with no Doppler effect. Another consequence of the absence of Doppler effect is the fact that the transit width is directly determined by the beam radius $w$, and not anymore the waist $w_0$.

*Selection of slow molecules*

To further improve the resolution, finding a way to enhance the relative contribution of the slowest particles (those that have the longest transit-time) to the saturation signal seems appealing[22]. In the absence of general methods for cooling molecules (as compared to atoms) an alternative is to extract a signal coming from the slowest molecules of the Maxwell-Boltzmann distribution. A given (low) pressure determines a characteristic transverse speed $v_\perp^0$ at which statistically only one collision occurs during the transit time of a particle through the laser beam. $v_\perp^0$ is of the order of $\gamma w_0$. Molecules of transverse speed $v_\perp < v_\perp^0$ (called *slow molecules* from now on) are in the collisional regime. They all experience the same coherent interaction time $\tau_c = 1/\gamma$ with the laser field which results in a homogeneous (constant amplitude and width) signal for this speed domain. One can easily choose the laser field which optimizes the signal for every particle of this velocity class (for instance $\Omega_R/\gamma = \pi/2$ for saturation spectroscopy, see above). Molecules of transverse speed $v_\perp > v_\perp^0$ (*fast molecules*) are in the free-flight regime. The interaction time is determined by the transit time $\sim w_0/v_\perp$ and decreases as well as the absorption probability and in turn the signal amplitude: this is the key condition that makes selection of slow molecules beneficial. However, the transverse speed $v_\perp$ distribution is equal to $\frac{2v_\perp}{u^2} e^{-(v_\perp/u)^2}$. This curve vanishes at $v_\perp = 0$ and reaches a maximum at $v_\perp = u/\sqrt{2}$. To reach the highest resolutions, one has to choose $v_\perp \ll u$. From the speed distribution, one sees that the selection is not so efficient as fast molecules will contribute significantly – with signal widths $\sim v_\perp/w_0$ – to the overall signal. Fortunately, the slow molecule selection can be enhanced by detecting successive derivatives of the line shape. After each derivation, the contribution of fast molecules is reduced by a factor $v_\perp^0/v_\perp$, (widths ratio) with respect to the homogeneous contribution of slow molecules. Experimentally, a low-frequency modulation *f* optimized for the detection of the narrowest signals from slow molecules of width $\gamma$ is applied. First- and second-harmonic detection enables to recover the first and second line shape derivatives. Expressions for the global line widths are summarized in Table 1 for both saturated absorption and Doppler-free two-photon spectroscopy[23,24]:

Table 1: Dependence of the line width (half-width at half-maximum in angular frequency) of the direct signal or the modulated signal after first- and second-harmonic detection for both saturated absorption and Doppler-free two-photon spectroscopy. Modulation depth is assumed to be small compared to $\gamma$.

|  | Saturated absorption | Doppler-free two-photon |
| --- | --- | --- |
| Non-derived line | $1.51 \sqrt{\gamma u/w_0}$ | $\frac{\ln 2}{2} \frac{u}{w}$ |
| First derivative | $1.44\, \gamma$ | $\sqrt{\gamma u/2\pi w}$ |
| Second derivative | $0.63\, \gamma$ | $\gamma/2$ |

Concerning the amplitude of the overall signal, let us take as a reference the amplitude obtained when the collisional width and transit width are of the same order for a transverse speed equal to $u$. It can easily be shown that, when reducing the pressure and $\gamma$ accordingly, the signal in the regime of selection of slow molecules is reduced by a factor $\gamma w_0/u^4$ (resp. $\gamma w/u^3$) in sub-Doppler saturated (resp. Doppler-free two-photon) absorption spectroscopy. The difference comes from the absence of selection of longitudinal velocities in the latter case. As a result, although the selection of slow molecules is an elegant method which leads to a gain in resolution by a factor of the order of $u/\gamma w_0$, which depends on the working pressure, the signal-to-noise ratio decreases very rapidly which prevented a wide use of this method. It has mainly been limited to proofs-of-principle or to experiments (such as the observation of ultra-narrow hyperfine structures) for which resolution is a critical requirement. As an example, ~100 Hz resolution ($3\times10^{-12}$ in relative value) was demonstrated in our group by slow-molecule detection in both sub-Doppler saturation spectroscopy[23] and Doppler-free two-photon spectroscopy[24] in the 18 m-long cell.

*Ramsey fringes*

The drastic reduction of the signal resulting from trying to increase the resolution limits the interest of methods based on the selection of slow molecules in a vapour cell. An alternative way to obtain higher resolutions is the famous method of separated fields first demonstrated by Ramsey[25]. This method can be combined with a sub-Doppler spectroscopic technique, such as saturated absorption or Doppler-free two-photon spectroscopy and allows reaching the highest resolutions in the optical domain. The general principle on which are based for example caesium beam clocks consists in a first field zone that addresses a transition in order to create a dipole which subsequently precesses freely at the molecular eigenfrequency. A second field interrogates the dipole and produces a population in the upper or lower state depending on the relative phase between the light and the dipole. This leads to a sinusoidal signal versus frequency around the resonance called Ramsey fringes. The obtained contrast is maximum with $\pi/2$ pulses in both zones. The resolution is proportional to the time of flight between the two field zones. Practically, this method is used to probe a molecular beam that crosses the Ramsey zones. In our simple explanation, we have omitted the Doppler effect. This is valid if the Ramsey method is combined with two-photon absorption spectroscopy, which is indeed Doppler-free, in which case the two field zones consist in two standing waves. For saturated absorption however, the basic configuration consists in two parallel travelling waves followed by two other anti-parallel waves with the same separation. The dipole which precesses between the two first zones accumulates a Doppler phase shift which will be exactly cancelled during the free flight between the two last zones. The parallelism between zones is thus crucial which usually implies the use of a single large corner cube or a cat's eye collecting all four beams. The resolution (and finally the interest of the method when the highest resolution is needed) is in turn limited by the optics size. By contrast, for Doppler-free two-photon Ramsey interferometry parallelism of the two standing waves is not required. Besides, the whole set of molecules in the lower level of the transition that crosses the two standing waves contribute to the Ramsey signal while the resolution is increased by the factor $L/w$ compared to conventional two-photon spectroscopy, where $L$ is the distance between zones. This makes Doppler-free two-photon Ramsey interferometry very attractive: simple geometrical design and contribution of all molecules. In addition, a molecular beam setup makes it easy to detect the molecular signal in a third zone, decoupled from the excitation zones, which enables the independent optimization of the detection laser power and thus signal-to-noise-ratio.

This latter technique allowed us to demonstrate a resolution of 100 Hz on a two-photon vibrational transition of $SF_6$ which led to a relative accuracy of $2\times10^{-14}$ of the optical frequency[9]. This is the strongest argument that led us to recently favour this spectroscopic method for the parity violation project.

*Overview of experimental results*

We conclude this section by summarizing our best experimental results implementing the various spectroscopic techniques considered above. For saturated absorption in the regime of selection of slow molecules[23], we recorded the first derivative lineshape of the $^{192}OsO_4$ P(39)$A_1^3$(-) line with a natural isotopic abundance of 40%, in the large cell (18 m long, total path length of 108 m, waist of 3.5 cm) and we obtained a peak-to-peak width of 220 Hz with a S/N$_{1Hz}$ of 6. For Doppler-free two-photon spectroscopy of slow molecules[24], we recorded the second derivative lineshape of the P(4)E$^0$ line of SF$_6$ in a Fabry-Perot of 3 m long, with a waist of 0.5 cm and we obtained a half-width at half-maximum (HWHM) of 280 Hz and a S/N$_{1Hz}$ of 2. Both results were outperformed by the two-photon Ramsey fringes[9], recording the same P(4)E$^0$ line on a supersonic beam of pure SF$_6$ with a distance between zones of 1 meter. We obtained a half-period of 100 Hz with a S/N$_{1Hz}$ of 30. The linear absorption spectroscopy performed on the $\nu_2$ saQ(6,3) rovibrational line of ammonia in an experiment aiming to measure the Boltzmann constant[7] led to a typical Doppler HWHM of 41 MHz. This poor resolution cannot be compensated by the very best S/N$_{1Hz}$ of 15000 to compete with the various sub-Doppler methods. Finally, when precision is concerned, the figure of merit $\frac{\Delta}{S/N_{1Hz}}$, where $\Delta$ is the line width, takes the following values (expressed in Hz) of 37, 140, 3 and 2700 for the respective experiments of saturated absorption, Doppler-free-two photon spectroscopy, two photon Ramsey fringes and linear absorption presented above.

**Laser frequency control**

As highlighted above, the invention of the laser enables to build spectrometers for which the light source did not limit the experimental resolution. In this Section we focus on CO$_2$ lasers which are commonly used for high-resolution rovibrational spectroscopy between 9 and 11 μm, and details the developments made at LPL over the past years of high resolution spectrometers based on stabilized CO$_2$ lasers. Lasers filled with the most abundant isotopic species $^{12}$C$^{16}$O$_2$ exhibit two emission bands centered at 9.4 and 10.4 μm, with emission lines found every 30 to 50 GHz. At LPL, we use home-made CO$_2$ lasers which have been carefully designed to optimize the passive stability[26]. They show a frequency drift of a few Hz/s. From the beat-note signal between two identical lasers, we deduced a free-running laser line width of about 1 kHz[26]. This makes possible rovibrational spectroscopy at a resolution of ~3×10$^{-8}$. It however requires to overcome the poor tunability of our $CO_2$ lasers, limited to ~100 MHz (or a few hundreds of MHz by shifting the frequency using acousto-optic modulators) around each rovibrational emission lines. We have thus built a broadband electro-optic modulator (EOM)[27] which enables to generate sidebands between 8 and 18 GHz allowing to cover a significant part of the 9-11 μm spectral range.

Ultra-high resolution molecular spectroscopy requires very stable lasers with very narrow line widths. To improve the long-term stability and reproducibility of our $CO_2$ lasers, we stabilize them onto molecular absorption lines, which constitute frequency references. To simultaneously enhance the short-term stability and the line width, the molecular line is detected in transmission of a Fabry-Perot cavity. In doing so, the signal-to-noise ratio increases proportionally to the cavity finesse, enabling a better frequency noise rejection[26,28]. Several molecules can be used for the $CO_2$ laser stabilization, among which $CO_2$[29,30] itself. The best stabilities were demonstrated with $SF_6$[31] and $OsO_4$[32,33] (using naturally abundant isotopic species, such as $^{190}OsO_4$ and $^{192}OsO_4$, presenting rovibrational lines with no hyperfine structure). Both molecules have rovibrational bands coinciding with the 10 μm $CO_2$ emission band[34]. $OsO_4$ was extensively studied since its $\nu_3$ band spectrum overlaps with the P(24) - R(26) $CO_2$ laser lines spectral region. A frequency grid of the strongest $OsO_4$ lines in coincidence with $CO_2$ laser lines has been established[35]. These lines have served for 20 years as secondary frequency standards in the mid-infrared region.

When $OsO_4$ is chosen as reference, the $CO_2$ laser is stabilized on a saturation signal detected in transmission of a Fabry-Perot cavity filled with $OsO_4$[32,33]. A broadband CdTe EOM[27], a key element of the stabilization scheme, generates sidebands, one of which is brought into resonance with the molecular line. Efficient and pure frequency modulations can then be applied to the sidebands simply by modulating the frequency of the synthesizer which drives the EOM. A first servo-loop locks the cavity resonance to the sideband frequency while a second locks the EOM sideband to the molecular line. Typical conditions for the detection of the $OsO_4$ saturated absorption signal are a pressure of 0.04 Pa and a laser power of 50 μW inside the cavity. In this regime, the third harmonic of the molecular signal has a peak-to-peak line width of about 20 kHz and a signal-to-noise ratio of 500 in a bandwidth of 1 kHz. Two equivalent and independent systems have been developed to characterize the frequency stability. By stabilizing two lasers onto the same strong P(46) $OsO_4$ line in coincidence with the P(14) $CO_2$ laser line we obtain an Allan deviation of $3.5\times10^{-14}$ $\tau^{-1/2}$ for integration times between 1 and 100 s, and a reproducibility of a few tens of Hz[33]. The associated laser line width of about 10 Hz makes it possible to do spectroscopy at resolutions in the $10^{-10}$ range.

The R(47) $A_2$ two-photon resonance of the $2\nu_3$ band of $SF_6$ was also used for the stabilization of the $CO_2$ laser. This is an attractive alternative to using $OsO_4$ because gaseous $SF_6$ is less reactive than $OsO_4$. Furthermore, all molecules contribute to this narrow two-photon resonance, whatever their speed, and for this particular line, the excitation probability is quite high due to the 16 MHz very small detuning of the intermediate level of the two-photon transition. Stability performances are similar to that obtained with $OsO_4$[31] but best stabilities can be maintained for significantly longer times[b].

$SF_6$ and $OsO_4$ do not show absorption in coincidence with the 9.4 µm band of the $CO_2$ laser. In this spectral range, stabilization on transitions of $CO_2$[29,30], $CHFClBr$[1] and $HCOOH$[36] was demonstrated with stability performances slightly degraded by the weaker absorption signal. An Allan deviation of $6\times10^{-13}$ $\tau^{-1/2}$ between 1 s and 100 s was for instance demonstrated using the $CO_2$ P(22) resonance, with a reproducibility of about 50 Hz[30].

The above stabilization techniques enable to reach excellent performances in the mid-infrared domain. However the obtained frequency reproducibility and accuracy are limited by long-term frequency fluctuations due to small variations in experimental parameters such as pressure or power, which induce line shifts or deformations[37]. Overcoming these fluctuations requires a better frequency reference which could for instance be provided by primary standards such as Cs clocks. Advanced primary standards are complex devices mainly developed in National Metrological Institutes (NMI). The French NMI (LNE-SYRTE) develops cold atoms based frequency standards in the microwave domain. Its Cs atomic fountain has a frequency stability of $1.6 \times 10^{-14}$ $\tau^{-1/2}$ and a demonstrated accuracy of a few $10^{-16}$ [38]. A local reference signal at 100 MHz[39] was developed with a stability provided by a H-Maser and an accuracy controlled with Cs fountains. The typical accuracy of the 100-MHz signal is below $10^{-14}$.

---

[b] $OsO_4$ being very reactive, the absorption cell has to be pumped and filled again after a few hours in order to maintain best stabilities. On the contrary, when using $SF_6$, the cell is filled only once a day.

Following what was done between SYRTE and Laboratoire Kastler-Brossel in Paris[40], we set up an optical fiber link between LPL and LNE-SYRTE in order to transmit the 100-MHz reference frequency signal to LPL[41]. The 43-km long optical link is subject to propagation noise due to thermal and acoustical fluctuations. The Allan deviation of frequency fluctuations added by the link is a few $10^{-14}$ at 1 s of measurement time and a few $10^{-15}$ after 1000 s which may degrade the spectral purity of the metrological signal. The propagation noise is thus actively compensated using the round-trip method: in the distant laboratory, the transferred signal is re-injected in the same fiber. This enables to measure the round-trip added noise in the initial laboratory and to correct for it in such a way that fluctuations are cancelled in the distant laboratory. When compensated, residual fluctuations are at the level of $10^{-14}$ at 1 s and $10^{-16}$ at 1000 s. Such performances enable to transmit the LNE-SYRTE frequency reference without any significant degradation[42].

Its frequency stability has then to be transferred to the mid-infrared spectral domain. Since the early 2000, optical frequency combs provided by femtosecond (fs) lasers have been largely developed for frequency comparison between the radiofrequency and the visible domain[43]. We adapted this technique to the measurement and control of the $CO_2$ laser frequency[41]. Using sum-frequency generation in a non-linear crystal of $AgGaS_2$, one can generate a beat-note between the $CO_2$ laser frequency and a very high harmonic of the repetition rate of the fs laser[31]. This beat-note is used to phase-lock the comb repetition rate to a sub-harmonic of the $CO_2$ laser frequency while the repetition rate is independently controlled with the reference signal from LNE-SYRTE. This technique enables to measure and control the $CO_2$ laser frequency with the primary standards of LNE-SYRTE. It was demonstrated using an $OsO_4$-stabilised $CO_2$ laser allowing to control its long-term frequency fluctuations and led to an accuracy of the order of 0.3 Hz, *i.e.* $10^{-14}$ in fractional value[9].

Even though at the state-of-the-art, the performances and reliability of this $CO_2$ laser stabilization scheme can be further improved. In the last ten years ultra-stable optical fiber links have been successfully developed enabling precise and accurate transfer of optical frequencies around 1.55 μm, with fractional frequency stability in the range of $10^{-18}$ after only 3 hours of measurement[44]. Moreover laser stabilization techniques have progressed in such a way[45] that near-infrared lasers can be stabilized on ultra-stable cavities leading to stabilities below $10^{-15}$ at 1 s measurement time[44]. Fiber frequency combs at 1.55 μm are now commercially available and easy to reference. Finally optical clocks are also progressing and are soon expected to reach accuracies in the $10^{-17}$ range[46]. Thus all ingredients are now available to directly stabilize the $CO_2$ laser frequency to a remote near-infrared optical reference without the need for a molecular reference and we have started to work in this direction. Moreover this technique could be extended to Quantum Cascade Lasers[47] which have a much larger tunability and achievable wavelengths covering the whole mid-infrared region. This will make it possible to perform accurate high-resolution spectroscopy on a wide range of molecular species.

**Parity violation and chiral molecules**

Any experimental attempt to measure parity violation in chiral molecules is inherently interdisciplinary, involving theoretical quantum chemistry, chemical synthesis and precision spectroscopy. A number of experimental techniques have been proposed for the observation of PV in chiral molecules, including rotational[48], rovibrational[17], Mössbauer[49] and NMR[50] spectroscopy, as well as crystallization[51] and solubility[52] experiments, optical activity measurements[53] or tunnelling dynamics of chiral molecules[54,55]. However, to our knowledge, apart from us only Martin Quack's group at ETH Zurich is currently pursuing an experimental effort, based on a proposal published in 1986 by Martin Quack himself, in which the PV energy difference is directly measured by probing the tunnelling dynamics of a molecule with a chiral ground state and an achiral electronic excited state[54].

We have chosen to follow another approach based on Letokhov's proposal of 1975 to observe PV effects in chiral molecules as a shift $\Delta \nu_{PV} = \nu_L - \nu_R$ in the rovibrational frequencies $\nu_L$ and $\nu_R$ of the same transition of left and right enantiomers[17]. Letokhov's group subsequently searched for such splittings in the spectrum of racemic CHFClBr by laser sub-Doppler absorption spectroscopy[56]. This approach was shortly after (in 1977) attempted on separated enantiomers of camphor[57]. From 1995, a third Letokhov-type experiment on enantioenriched samples of CHFClBr molecules was carried out at LPL[1,58]. This experiment used a $CO_2$ laser-based spectrometer to probe a hyperfine component of the C–F stretch at $\nu \sim 30$ THz (10 μm or 1000 cm$^{-1}$) via saturated absorption laser spectroscopy. The spectrum of each enantiomer was simultaneously recorded in separate Fabry–Perot cavities. A experimental sensitivity of $2 \times 10^{-13}$ was attained, setting an upper limit of $\Delta \nu_{PV} < 8$ Hz, limited by residual differential pressure shifts induced by impurities in the samples[58]. Shortly after, however, theoretical studies concluded that the PV vibrational shift for the C–F stretch of CHFClBr is on the order of 2.4 mHz[59], corresponding to $\Delta \nu_{PV}/\nu \sim 8 \times 10^{-17}$. This effort showed that further progress would require both an improved experiment (in particular, where collisional effects are negligible) and a conscientious effort to find better candidate molecules.

We are now working on the development of a second generation $CO_2$ laser spectroscopy experiment based on 2-photon Ramsey interferometry of chiral molecules in a continuous supersonic jet (see Section Ramsey fringes), aiming for an experimental sensitivity of a few 0.01 Hz[3,60,61]. We have also launched a collaboration in which the search for suitable chiral molecules is guided by relativistic molecular calculations by Trond Saue and co-workers and the synthesis of candidate molecules is directed by Jeanne Crassous. Thérèse Huet and Pierre Asselin are responsible for the spectroscopic characterization of candidate molecules in the microwave and infrared domain respectively. As a result of this collaboration, the focus has shifted to studying chiral complexes of heavy metals for which theoretical studies[c] have recently predicted PV shifts on the order of 1 Hz[63]. In view of current experimental conditions, several criteria have been outlined for the appropriate chiral molecule for a PV test. The ideal candidate should:

(1) show a large PV vibrational frequency difference $\Delta\nu_{PV}$ of an intense fundamental transition within the $CO_2$ laser operating range (850-1120 cm$^{-1}$);

(2) be available in large enantiomeric excess or, ideally, in enantiopure form;

(3) be available at gram-scale;

(4) have high enough volatility or sublimate without decomposition to allow gas phase studies;

(5) have a suitable 2-photon transition joining a state in the fundamental vibrational level, v = 0 to one in the v = 2 level;

(6) keep the structure as simple as possible so as to maintain a favourable partition function and facilitate the spectroscopy.

J. Crassous and co-workers have first synthesized two classes of rhenium complexes in enantiopure form (see Figure 1): oxorhenium compounds such as **1**, bearing hydrotris(1-pyrazolyl)borate (Tp) ligand and a chiral bidentate ligand[64] and '3+1' oxorhenium complexes bearing a tridentate sulfurated ligand and a monodentate halogen (such as in **2**) or a chalcogenated ligand [65,66]. However, these candidate molecules, which are solid at room temperature, have been found not to satisfy the volatility requirement 4), needed when using our existing supersonic beam apparatus[3]. Chemists then focused on chiral analogues of methyltrioxorhenium (MTO, **3** on Figure 1), a rhenium complex with somewhat better sublimation characteristics[d]. Complexes **4** has recently been synthesized[68] and show similar volatility to MTO. For all the above mentioned oxorhenium complexes, the oxo ligand gives an intense band around 1000 cm$^{-1}$, corresponding to its Re=O stretch. Calculations by T. Saue and co-workers indicate that PV shifts $\Delta\nu_{PV}$ in complexes such as **2** and **4** can reach several hundreds of millihertz[66,68], about an order of magnitude above the expected sensitivity of a differential measurement using 2-photon Ramsey interferometry of a molecular beam.

---

[c] There has been quite a significant activity in the quantum chemical community in the past years with important contributions from the groups of M. Quack, R. Berger, P. Lazzeretti, P. Schwerdtfeger, P. Manninen and T. Saue that have been summarized in recent reviews[61][62].

[d] Note that isotopically chiral MTO (CH$_3$Re$^{16}$O$^{17}$O$^{18}$O) had already been suggested by Martin Quack[67].

At LPL, we are studying MTO, the achiral parent molecule, as a preliminary step towards making an ultra-sensitive PV measurement. Precise measurements on such large and complex molecules are rare. We needed to gain both insight on the apparatus needed for such a measurement, and experience with conducting experiments on spectroscopically novel species. We thus conducted high resolution spectroscopy of MTO, in a cell at room temperature and in a cold supersonic beam. The ultra-high resolution spectrometer used is based on the combination of two $CO_2$ lasers, the first locked to an $OsO_4$ rovibrational line[33] (see Section Laser frequency control), the other, whose beam is used to probe MTO, phase-locked to the first. Acousto-optic modulators allow tuning over a few hundreds of megahertz around each $CO_2$ laser line[69].

Results of spectroscopy in a cell are presented first. Saturated absorption spectra of room temperature MTO were recorded in a 58-cm long cell, around the *R*(18), *R*(20), *R*(22) and *R*(24) $CO_2$ laser lines[69]. Figure 2 shows a spectrum recorded over 30 MHz around the *R*(20) $CO_2$ laser line revealing a dense hyperfine structure. Linewidths (full width at half maximum) below 100 kHz and central frequency accuracies of a few kilohertz were obtained. Collaborative work combining our spectra with data from Fourier transform microwave and infrared spectroscopy enabled the assignment of rovibrational lines, some of them with their resolved hyperfine structure (see Figure 2)[69]. A set of spectroscopic parameters in the ground and first excited state, including hyperfine structure constants, was obtained for the antisymmetric Re=O stretching mode of the $^{187}$Re MTO isotopologue. This result validates the experimental approach to be followed with chiral derivatives of MTO in order to identify the best candidate line to be studied for PV observation. We move on to the results from the supersonic beam. At LPL, we had, at our disposal, a molecular beam setup which was previously used in an ultra-high resolution two-photon Ramsey-interferometry experiment on SF$_6$ [9] (see Section Ramsey fringes). We added an oven to the continuous supersonic beam source to allow MTO to be sublimated (by heating it to near 100°C, above which its starts to decompose) and seeded in a carrier gas. In our apparatus, supersonic expansion occurs through a circular nozzle in a first chamber ($10^{-5}$–$10^{-4}$ mbar under working conditions) separated from the second one ($10^{-6}$–$10^{-5}$ mbar under working conditions) by a skimmer. Extensive studies of the MTO-seeded jet characteristics as a function of reservoir temperature, backing He pressure, nozzle-to-skimmer distance, nozzle and skimmer diameter have been performed *via* time-of-flight (TOF) experiments. A rotating slotted disk was used to chop the molecular jet and an ionisation gauge placed on the jet axis enabled to record TOF[70]. Density-weighted longitudinal velocity ($v_{long}$) distribution[e] written as

$$f(v_{long}) \propto v_{long}^2 e^{-\left(\frac{v_{long}-V}{\Delta v_{long}}\right)^2},$$

was extracted from a fit of the TOF, carefully considering the finite opening time of the chopper and the finite dimensions of the gauge. The MTO molar fraction $x$ was obtained from the mean velocity $V$ [71]:

$$V = \sqrt{2k_B T_{nozzle} \frac{\frac{\gamma_{MTO}}{\gamma_{MTO}-1}x + \frac{\gamma_{He}}{\gamma_{He}-1}(1-x)}{m_{MTO}x + m_{He}(1-x)}}$$

where $k_B$ is the Boltzmann constant, $T_{nozzle}$ the nozzle temperature, $\gamma_S$ and $m_S$ the Poisson coefficient and molecular mass of species S. Considering the low MTO molar fractions obtained (< 10%, see Table I), the translational longitudinal temperature $T_{long} = m_{He}\Delta v_{long}^2/2k_B$ was calculated from the velocity spread $\Delta v_{long}$, assuming the TOF signal to be dominated by the He contribution[f]. Table I summarizes the results obtained for different pairs of nozzle and skimmer diameters and corresponding He backing pressure that maximizes $x$.

---

[e] $v_{long}$ means the velocity component along the molecular beam axis, which is usually almost perpendicular to the laser beam and thus close to $v_\perp$, the speed transverse to the laser beam axis introduced in Section Molecular ultra-high resolution spectroscopy.

Table I: MTO-seeded jet characteristics (deduced from TOF experiments) for different pairs of nozzle and skimmer diameters ($d_{nozzle}/d_{skimmer}$) and corresponding He backing pressure ($P_{He}$) that maximizes the MTO molar fraction $x$. The MTO partial pressure in the gas line $P_{MTO}$ is calculated as $P_{He} \times x$. The oven and nozzle temperatures were respectively 80 and 100°C.

| $d_{nozzle}/d_{skimmer}$ | 50 μm/750 μm[g] | 100 μm/750 μm | 200 μm/2 mm | 300 μm/2 mm |
|---|---|---|---|---|
| $P_{He}$ (mbar) | 1000 | 200 | 150 | 100 |
| $V$ (m/s) | 1386 ± 19 | 1007.1 ± 0.5[h] | 823.5 ± 0.8[h] | 865.7 ± 1.0[h] |
| $\Delta v_{long}$ (m/s) | - | 114.4 ± 1.3[h] | 139.9 ± 1.5[h] | 157.2 ± 1.6[h] |
| $x$ (%) | 1.8 ± 0.2 | 5.2 ± 0.5[i] | 9.5 ± 1.3[i] | 8.2 ± 1.0[i] |
| $P_{MTO}$ (mbar) | 18 ± 2 | 10.5 ± 0.9[i] | 14.2 ± 2.0[i] | 8.2 ± 1.0[i] |
| $T_{long}$ (K) | - | 3.1 ± 0.1[h] | 4.7 ± 0.1[h] | 5.9 ± 0.2[h] |

[g] in these experimental conditions, only the mean velocity $V$ was roughly measured.

[h] uncertainty given by the TOF fit.

[i] uncertainty is dominated by that on $\gamma_{MTO}$, the unknown effective MTO Poisson coefficient during thesupersonic expansion, assumed to be between 1.11 and 1.33.

Linear absorption spectroscopy carried out on this setup led to the first spectra of a supersonic beam of MTO with a ~1 MHz resolution. The chopper described above was used to induce a ~1 kHz modulation of the absorption detected with a lock-in amplifier. Line shapes are Doppler broadened owing to the molecular jet divergence. First spectra with a rather poor signal-to-noise ratio were recorded in a double-pass configuration[69]. We were able to increase the signal by a factor of ~7 after a multi-pass cell was installed inside the jet vacuum chamber, allowing nine parallel laser beam passes (one every 2 cm) to probe the molecular jet perpendicularly (the laser beam was retroreflected leading to an 18-pass configuration). This paved the way for a systematic study of the different experimental parameter in view of optimizing the spectroscopic signal. As illustrated on Figure 3, reducing the nozzle diameter from 200 µm to 100 µm induced a temperature decrease (from ~10 K to ~1 K) leading to a simplified spectrum, with no significant loss in signal-to-noise ratio. Even if lowering the total flux by a factor of roughly 4, reducing the nozzle diameter made it possible to increase He backing pressure and in turn decrease the temperature further thus maximizing the population of low $J$ rotational quantum number levels. In agreement with spectra simulated from the set of parameters obtained in the Re=O stretching mode analysis mentioned above[69], signal is preserved in zones where the lowest $J$ transitions are expected and disappears in the highest $J$ transition zones (see Figure 3).

---

[f] Note that this probably leads to slightly underestimated $T_{long}$.

The next step on the jet experiment is the observation of saturated absorption features. A careful comparison between linear and saturated absorption data obtained in the cell led us to conclude that further improvements in signal-to-noise ratio is required in order to observe saturated absorption in a supersonic beam of MTO. To that purpose, the installation of an in-vacuum high-finesse (~1800) Fabry-Perot cavity is under progress. This should make it possible to observe saturated absorption and Ramsey fringes in a jet of MTO, and should lead to the demonstration of both ultra-high resolution spectroscopy and coherent manipulation of oxorhenium complexes.

**Conclusion**

We have strived over the years to observe parity violation in chiral molecules. This paper presented an overview of the spectroscopic methods and apparatus we have developed in the hope of reaching this goal. We find that applying Ramsey interferometry on a continuous supersonic molecular beam of enantioseparated chiral molecules appears, at the moment, to be the most promising avenue towards observing PV in rovibrational transitions. This, however, may change in the near future. Molecular beam source technology is undergoing drastic changes thanks to the development of buffer gas cooled beams[72]. Using these beams as a starting point, the first demonstrations of trapped, polyatomic, relatively complex molecules are emerging[73]. We are currently considering the implementation of these cold, slow beams for the next generation PV experiment at LPL. On the laser stabilization side, we started working on stabilizing quantum cascade lasers (QCL) in the hope of eventually replacing our $CO_2$ lasers. QCLs being far more tunable, we hope to use these lasers to probe many more molecular species.

The observation of PV in chiral molecules has been a long-standing dream for both Martin Quack, and us. On this occasion, we wish Prof. Quack a happy birthday and good luck in this challenging endeavour.

**Acknowledgments**

The authors acknowledge financial support from Centre National de la Recherche Scientifique (CNRS), Région Île-de-France and Université Paris 13. This work is part of the project NCPCHEM 2010 BLAN 724 3 funded by the Agence Nationale de la Recherche (ANR, France). We would like to thank its co-ordinator Trond Saue and participants (Jeanne Crassous, Nidal Saleh, Laure Guy, Radovan Bast, Thérèse R. Huet, Pierre Asselin, Pascale Soulard,…) for valuable discussions. O. Lopez is gratefully acknowledged for his contribution to this project.

**Figure Captions**

Figure 1: Classes of chiral and achiral oxorhenium complexes considered for the experiment.

Figure 2: Saturated absorption spectrum of MTO in a cell at 300 K detected after frequency modulation at 5 kHz (200 kHz modulation depth, applied on piezoelectric transducers that control the laser cavity length) and 2$^{nd}$ harmonic detection in order to strongly flatten the baseline induced by the laser gain curve. Vertical coordinate is the signal amplitude in arbitrary units. Experimental conditions: 1 point every 10 kHz, 200 ms of integration time per point, 0.002 mbar of MTO, pump (resp. retroreflected probe) beam power of 95 µW (resp. 12 µW), 5 to 10 mm beam waist. The assigned $\Delta F = 0$ six most intense hyperfine components associated with the $^{R}Q(J=20, K=0)$ line of the antisymmetric Re=O stretching mode of $^{187}$Re MTO are labelled with their $F$ quantum number[69].

Figure 3: Linear absorption spectroscopy of a MTO-seeded supersonic jet, in the vicinity the R(20) $CO_2$ laser line frequency (corresponding to 0 MHz on the bottom axis). Lower pink curve: 200 µm (resp. 2 mm) nozzle (resp. skimmer) diameter, $P_{He}$ ~150 mbar, $T_{oven}$ ~80°C, $T_{long} \geq 5$ K, 7.5 mm nozzle-to-skimmer distance. Upper blue curve: 100 µm (resp. 750 µm) nozzle (resp. skimmer) diameter, $P_{He}$ ~1 bar, $T_{oven}$ ~90°C, $T_{long}$ ~ 1 K, 9 mm nozzle-to-skimmer distance. Other experimental conditions: 1 point every 100 kHz, ~1 s of integration time per point, ~0.5 µW of laser power in each of the 18 passes, ~5 mm beam waist. The inset is a zoom on a single line (1 point every 400 kHz, 4 s of integration time per point, signal-to-noise ratio ~10).

Figure 1

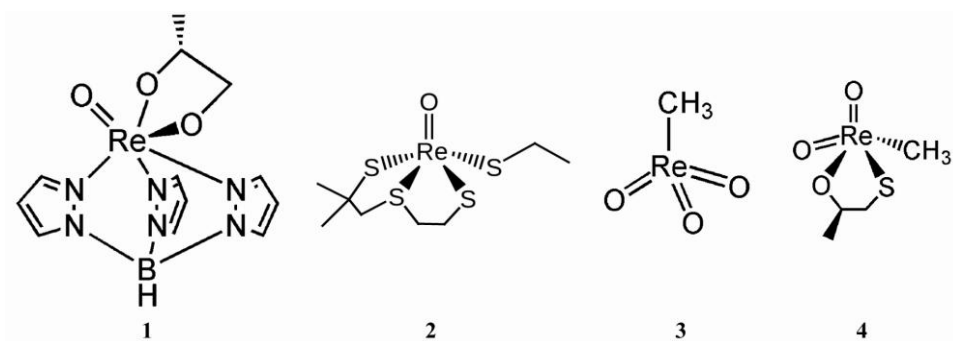

Figure 2

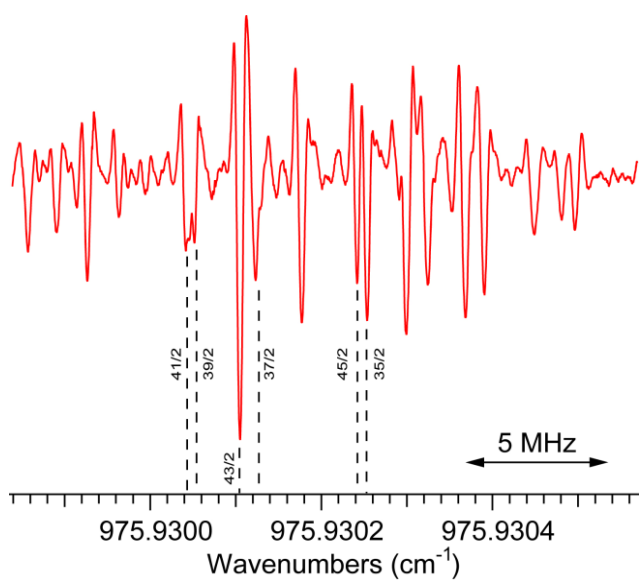

Figure 3

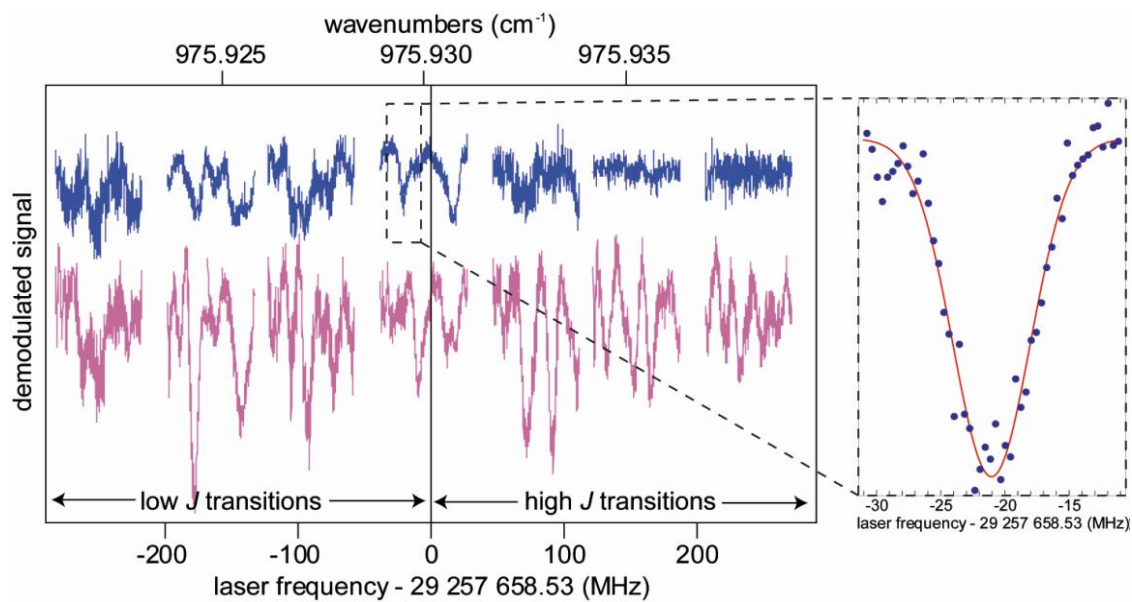